\documentclass{amsart}

\usepackage{amsmath,amssymb,amsfonts,amscd,amsthm,amsopn,epsfig}  
\usepackage{amsthm}
\usepackage{amssymb}
\usepackage{latexsym,verbatim}

\usepackage{mathrsfs}
\usepackage{enumerate}

\usepackage{tikz}\usetikzlibrary{backgrounds,matrix,arrows,shapes,decorations.markings}
\newcommand{\btp}{\begin{tikzpicture}[baseline=-5pt,scale=0.25,line width=0.7pt]}
\newcommand{\etp}{\end{tikzpicture}}

\pgfdeclarelayer{background}
\pgfdeclarelayer{foreground}
\pgfsetlayers{background,main,foreground}

\usepackage{hyperref}
\hypersetup{
    citecolor=blue,        
}

\usepackage{color}

\newtheorem{rmk}{Remark}[section]

\numberwithin{equation}{section}

\def\V={{{\bf\rm{V}}}}

\def\beq{\begin{equation}}
\def\eeq{\end{equation}}
\def\ben{\begin{eqnarray}}
\def\een{\end{eqnarray}}
\def\ba{\begin{array}}
\def\ea{\end{array}}

\def\lt{\left}
\def\rt{\right}

\newcommand\cW{{\mathcal W}} 
\newcommand\cH{{\mathcal H}}

\newcommand\CC{\mathbb C}

\newcommand\Om{|\Omega\rangle}
               
\newcommand{\so}{\scriptscriptstyle \rm I}
\newcommand{\st}{\scriptscriptstyle \rm I\hspace{-1pt}I}

\begin{document}
%

%
\title{Modified algebraic Bethe ansatz for XXZ chain on the segment
\\ - I - triangular cases
}
\author{S.~Belliard}
\address{ Laboratoire de Physique Th\'eorique et Mod\'elisation (CNRS UMR 8089), Universit\'e de Cergy-Pontoise, F-95302 Cergy-Pontoise, France }
\email{samuel.belliard@gmail.com}
\begin{abstract}
The modified algebraic Bethe ansatz, introduced by Cramp\'e and the author \cite{BC13}, is used to characterize the spectral problem of the Heisenberg XXZ spin-$\frac{1}{2}$  chain on the segment with lower and upper triangular boundaries. 
The eigenvalues and the eigenvectors are conjectured. 
They are characterized by a set of Bethe roots with cardinality equal to $N$ the length of the chain and which satisfies a set of Bethe equations with an additional term. 
The conjecture follows from exact results for small chains.  
We also present a factorized formula for the Bethe vectors of  the Heisenberg XXZ spin-$\frac{1}{2}$  chain on the segment with two upper triangular boundaries. 
\end{abstract}

\maketitle


\vskip -0.2cm



\vspace{5mm}

\section{Introduction}

Let us consider the Heisenberg XXZ  spin-$\frac{1}{2}$ chain on the segment given by the Hamiltonian
\ben\label{H}
&&H= \epsilon\, \sigma^{z}_1 +  \kappa^-\,\sigma^{-}_1  +  \kappa^+\,\sigma^{+}_1   +\sum_{k=1}^{N-1}\Big(\sigma^{x}_{k}\otimes \sigma^{x}_{k+1}+\sigma^{y}_{k}\otimes \sigma^{y}_{k+1} +\Delta\sigma^{z}_{k}\otimes \sigma^{z}_{k+1}\Big)  + \nu\,  \sigma^{z}_N +\tau^-\,  \sigma^{-}_N+\tau^+\,  \sigma^{+}_N,
\een
where $N$ is the length of the chain and $\{\sigma_i^j\}$, with $j\in\{x,y,z,+,-\}$, are the Pauli matrices\footnote{
 $
\sigma^{z}=\left(\begin{array}{cc}
       1 & 0\\
      0 & -1      \end{array}\right),\quad
\sigma^{+}=\left(\begin{array}{cc}
       0 & 1\\
      0 & 0      \end{array}\right),\quad
\sigma^{-}=\left(\begin{array}{cc}
       0 & 0\\
      1 & 0      \end{array}\right),\quad
      \sigma^{x}=\sigma^{+}+\sigma^{-}, \quad  \sigma^{y}=i(\sigma^{-}-\sigma^{+}) $.} 
that act non-trivially on the site $i$ of the quantum space $\cH= \otimes_{i=1}^N V_i$ with $V_i=\CC^2$. Here $\Delta=\frac{q+q^{-1}}{2}$, where $q$ is a generic parameter, denotes the anisotropy parameter and $\{ \epsilon,  \kappa^\pm\}\in \CC^3$ are the left boundary parameters and $\{ \nu,  \tau^\pm\}\in \CC^3$ are the right boundary parameters. 
This model is among the simplest open quantum integrable models on the lattice and finds applications in a wide range of domains such as condensed matter, high energy physics, out of equilibrium statistical physics, mathematical physics, etc.  

\vspace{0.2cm}

For diagonal boundaries, $\kappa^\pm=\tau^\pm=0$, the spectral problem of this model has been firstly characterized in \cite{Gau83,ABB87} by mean of the coordinate Bethe ansatz (CBA) introduced by Bethe \cite{Bet31}. 
The Bethe ansatz (BA) corresponds to parametrize the eigenvalues and eigenvectors by a set of parameters that satisfy a system of coupled equations, the so-called Bethe equations (BE). 
Then, the hidden spectrum generating  algebra of the model (\ref{H}) has been identified by Sklyanin \cite{Skly88} in terms of the reflection equation \cite{chered} and the associated reflection algebra. 
Roughly speaking, the hidden spectrum generating  algebra of quantum integrable lattice models is given by a quantum group for models on the circle and by a coideal sub-algebra of a quantum group
\footnote{a reflection algebra is a possible realization of a coideal sub-algebra of quantum group.}
for models on the segment. 
For the XXZ chain one considers the quantum algebra $U_q(\widehat{gl_2})$ and its coideal sub-algebras that allow one to construct conserved quantities and the Hamiltonian for the model on the circle and on the segment, respectively. 
The origin of these algebras is the so-called quantum inverse scattering method introduced by Faddeev school \cite{SFT}. 
In this framework, the algebraic BA (ABA) was performed by Sklyanin \cite{Skly88} for diagonal boundaries recovering the eigenvalues and BE found from the CBA and providing a factorized realization of the Bethe wave equation, the so-called Bethe vectors (BV). 
These BV are crucial to study the correlation functions performed in \cite{KKMNST}. 
For diagonal boundaries, the Hamiltonian has a $U(1)$ symmetry. Indeed, it commutes with the total spin operator $J^z=\frac{1}{2}\sum_{i=1}^N\sigma_i^z$. 
Thus the quantum space and the spectral problem decompose into the direct sum of the invariant $J^z$ subspaces $\cW_i$
\ben
\cH=\oplus_{i=0}^{N} \cW_i
\een 
with  $J^zv_i=\frac{N-2i}{2}v_i$ for all $v_i\in \cW_i$.
This condition is required to apply the usual BA for quantum integrable models on the segment and on the circle.  
It allows one to find a simple eigenvector of the Hamiltonian, the so-called highest weight vector (or reference state). 
This highest weight vector is the only vector of $\cW_0$ and others subspaces can be constructed from the action of the so-called creation operator that belong to the hidden spectrum generating  algebra. 
The BA is then applied independently to each subspace.  

\vspace{0.2cm}

For non-diagonal boundaries, the breaking of the $U(1)$ symmetry by the off-diagonal boundary terms, parametrized by $\{\tau^\pm,\kappa^\pm\}$, does not allow one to decompose the spectral problem. 
Thus the usual BA fails to provide the spectral problem for all range of the parameters although the hidden spectrum generating  algebra of the model is known and quantum integrability admitted.
More generally this is the case for a lot of quantum integrable models without $U(1)$ symmetry. 
Finding methods to consider the spectral problem of such models is an active field of research. 
Let us recall some results for the XXZ chain on the segment.  

By imposing constraints between right and left boundary parameters, the spectral problem has been characterized by mean of different BA: ABA \cite{gauge,YanZ07}, Analytical BA \cite{TQ,TQ2}, CBA \cite{CRS1}. 
These constraints correspond, up to some similarity transformation, to consider an auxiliary model with diagonal boundaries or  with a diagonal and a triangular boundaries \cite{Baj}. 
For the latter the CBA was applied in \cite{CRS1,CRS2}, the wave function involves linear superposition of $U(1)$ subspaces and the eigenvalues are the same than for two diagonal boundaries, see also \cite{MRM} for the XXX chain and by mean of the ABA. 
For two upper triangular boundaries, {i.e.} $\kappa^-=\tau^-=0$, the problem was considered in \cite{pimenta} (see also \cite{BCR12} for XXX chain). 
It also leads to the same eigenvalues than for the diagonal/upper triangular case but with new BV. They are linear superpositions of the diagonal/upper triangular BV.
For generic boundary parameters and for $q=e^{i\frac{\pi}{p}}$, root of unit, an analytical BA was proposed \cite{qroot}. The eigenvalues are parametrized by Bethe roots that satisfy a non-conventional BE with a lot of terms increasing with $p$. 

The first solution for generic parameters is due to an alternative approach to the BA, the so-called Onsager approach \cite{BK}, that takes its roots in the initial paper of Onsager on the two dimensional Ising model \cite{Ons44}, see \cite{Bas} for details. 
This approach is based on a new realization of the coideal sub-algebra of $U_q(\widehat{gl_2})$ called the q-Onsager algebra. In this approach, the characterization of the complete spectral problem
\footnote{the completeness of the spectrum follows from the representation theory of the q-Onsager algebra \cite{BK}}
 is given by the roots of the characteristic polynomial of a block tridiagonal matrix. 
At the same period another solution for the eigenvalues was proposed from functional approach \cite{Gal08}. In this case the spectral problem is characterized by some BE that have a non-conventional structure and remain to be explored.  

More recently the development of the quantum separation of variables (SoV), introduced by Sklyanin \cite{SklySoV1,SklySoV2}, has allowed to characterize the spectral problem of the inhomogeneous transfer matrix related
\footnote{The Hamiltonian (\ref{H}) is constructed from the homogeneous transfer matrix, thus the SoV does not characterize directly its spectrum.} 
 to the Hamiltonian
(\ref{H}) \cite{niccoli2,FKN}. An important point for the SoV  is that the completeness of the spectral problem follows by construction
, which is not the case for the BA where completeness is admitted from numerical checks. 
Another recent result is the off-diagonal Bethe ansatz (ODBA), introduced by Cao {\it et al.}  \cite{CYSW1}, that extends the analytical BA to all models without $U(1)$ symmetry in term  of "quite" conventional BE. The main feature consists in adding a new term in the eigenvalues and the BE
\footnote{ The ODBA allows one to consider different parametrization for the eigenvalues. Numerical check on XXX and XXZ chains on the segment shows that all of them give a full description of the spectral problem \cite{comXXXCao,Nep,CYSW1}.}, see also \cite{frahm} for this idea. 
The ODBA has been applied to many models (see references in \cite{CYSW4}) and in particular to the Hamiltonian (\ref{H}) \cite{CYSW3}.

The connection between the SoV approach, the eigenvalues and the BE with a new term from the ODBA
was given in \cite{KMN14}. In the reverse, the basis used in the SoV approach to construct the states allows one to retrieve the eigenstates in the ODBA   \cite{CYSW4}.  It shows the deep relation between BA and SoV. 
Indeed, the completeness of the BE solution, at least in inhomogeneous case, results from  the SoV; and in the reverse side, BA provides a regularization scheme for taking the homogeneous limit of SoV characterization of the spectrum. 

\vspace{0.2cm}

So an important step for the future developments of the BA and SoV characterizations of the spectrum problem of quantum integrable models is to construct the BV associated to models without $U(1)$ symmetry.    
A first result in this direction was for the Heisenberg XXX spin-$\frac{1}{2}$  chain on the segment with general boundaries. The BV was conjectured by Cramp\'e and the author \cite{BC13}. 
In principle, such BV exists for other models without $U(1)$ symmetry and thus can be at least conjectured from what we call modified algebraic Bethe ansatz (MABA).  
The MABA is independent of the inhomogeneous parameters used in the SoV and ODBA approaches and thus characterizes directly the full spectrum problem of the model. 

\vspace{0.2cm}
   
Here, we present the MABA for XXZ spin chain on the segment with upper and lower boundaries and we conjecture the BV and the eigenvalue. 
We construct the conjecture independently of the knowledge of the eigenvalue
\footnote{ The initial procedure \cite{BC13} uses the simplest parametrization of the eigenvalue presented in \cite{CYSW2,Nep}.}
. 
Thus the MABA allows one to conjecture the eigenvalues and the eigenvectors of the model. 
We also revisit the case with two upper boundaries \cite{pimenta} and present a factorized formula of the associated BV that provides an algebraic proof similar to the one of the usual ABA. 
Moreover it is an important intermediate step to understand the MABA.

\vspace{0.2cm}

The paper is organized as follow: 
in section \ref{RKRKform} we recall basic properties of the quantum group $U_q(\widehat {gl_2})$ in RLL realization and of its coideal sub-algebra in reflection algebra realization used to apply the ABA and MABA.
Then, in section \ref{S:UABA}, we recall the ABA that we apply for diagonal/diagonal and diagonal/upper triangular boundaries conditions.
In section \ref{S:MABAupup}, the case with two upper triangular boundaries is revisited and 
in section \ref{S:MABA} we give the modified algebraic Bethe ansatz for lower/upper triangular boundaries and conjecture eigenvalues and eigenvectors.  
The section \ref{S:smallN} is devoted to the construction of the conjecture for the small length cases $N=1,2$. 
Finally, in section \ref{S:Conc} we discuss the extension of the MABA to general boundary cases and some perspectives for the presented results.

\vspace{0.2cm}

 {\bf Notations :}
 We will use the notation $\bar u $ and $\# \bar u =a$ for the set of $a$ variables $\{u_1,u_2,\dots,u_a\}$. 
 If the element $u_i$ is removed, we denote $\bar u_i =\{u_1,u_2,\dots,u_{i-1},u_{i+1},\dots,u_a\}$. 
 For the product of functions or of commuting operators we use the convention
 \ben
 f(u,\bar u)=\prod_{i=1}^af(u,u_i), \quad \mathscr{B}(\bar u)=\prod_{k=1}^a\mathscr{B}(u_k).\nonumber
\een
We will also use the so-called auxiliary space framework. 
For any given matrix $A$ in $End(V)$ we denote $A_i$ the matrix that act non-trivially on $V_i$ in the multiple tensor product vectorial space $V_1 \otimes \dots \otimes V_m$. For any given matrix $B$ in $End(V\otimes V)$ we denote $B_{ij}$ the matrix that act non-trivially only on $V_i$ and $V_j$. Here we will always have $V=\CC^2$. 
All functions and commutation relations between operators used in the paper are gathered in the appendix \ref{App:Func}.
To consider the XXX limit and recover notations of \cite{BC13} one has to consider
\ben
u=e^{\hbar \lambda}, \quad q= e^{\hbar},  \quad \nu^\pm=\mp\frac{e^{\mp\hbar p}}{ e^{\hbar}- e^{-\hbar}},  \quad \kappa=\frac{\xi^-}{2( e^{\hbar}- e^{-\hbar})},  \quad \tilde \kappa=\frac{\xi^+}{2( e^{\hbar}- e^{-\hbar})},\nonumber
 \\
 \epsilon_\pm=\pm\frac{e^{\pm\hbar q}}{ e^{\hbar}- e^{-\hbar}},  \quad \tau=\frac{\eta^+}{2( e^{\hbar}- e^{-\hbar})},  \quad \tilde\tau=\frac{\eta^-}{2( e^{\hbar}- e^{-\hbar})} \quad \mbox{and to take the limit}\, \hbar \to 0.\nonumber
\een
 


\section{XXZ chain on the segment from reflection algebra \label{RKRKform}}

The Hamiltonian (\ref{H}) can be constructed from the reflection algebra (RKRK) following \cite{Skly88}. 
Here we recall basic properties of the quantum group $U_q(\widehat {gl_2})$ in RLL realization and of its coideal sub-algebra in RKRK realization. 
Then we recall the fundamental highest weight representation of $U_q(\widehat {gl_2})$ and give the action of its coideal generators on the highest weight vector when the right boundary is upper triangular. 

\subsection{Quantum group $U_q(\widehat {gl_2})$ in RLL realisation\label{s:RLL}}
Let us recall the symmetric\footnote{{\it i.e.} $R_{ab}(u)=R_{ba}(u)$.} trigonometric R-matrix 
\ben\label{R}
\qquad R_{ab}(u)=\lt(\begin{array}{cccc}b(qu)&0&0&0\\0&b(u)&1&0\\
0&1&b(u)&0\\0&0&0&b(qu)\end{array}\rt)=\lt(\begin{array}{cc}b\big(q^{\frac{1+\sigma_b^{z}}{2}}u\big)& \sigma_b^{-}\\ \sigma_b^{+} & 
b\big(q^{\frac{1-\sigma_b^{z}}{2}}u\big)\end{array}\rt)_a,\quad
b(u)=\frac{u-u^{-1}}{q-q^{-1}},
\een
solution of the quantum Yang-Baxter equation 
\ben\label{YB}
R_{ab}(u_a/u_b)R_{ac}(u_a/u_c)R_{bc}(u_b/u_c)
=R_{bc}(u_b/u_c)R_{ac}(u_a/u_c)R_{ab}(u_a/u_b).
\een 
The quantum algebra $U_q(\widehat{gl_2})$ can be realized by the (one row) quantum monodromy matrix 
\ben\label{L}
L(u)=\left(\begin{array}{c c}
      l_{11}(u) & l_{12}(u) \\
   l_{21}(u)  & l_{22}(u) 
      \end{array}
\right),
\een
with $l_{ij}(u)=\sum_{n=0}^\infty u^{-n} l^{(n)}_{ij}$ the generating functions of the generators of the algebra\footnote{In fact $L(u)$ contains only the positive Borel sub-algebra of $U_q(\widehat{gl_2})$ and one has to consider another monodromy matrix and a central element to have the full set of generators of this algebra.}. The quantum monodromy matrix satisfies the RLL relation
\ben\label{RLL}
R_{ab}(u/v)L_a(u)L_b(v)=L_b(v)L_a(u)R_{ab}(u/v).
\een
The center $Z$ of $U_q(\widehat{gl_2})$ is given by the quantum determinant of the monodromy matrix
%
%
\ben\label{LDet}
{\rm Det}_q\{L(u)\}=tr_{ab}(P_{ab}^{-}L_{a}(u)L_{b}(qu))=l_{11}(qu) l_{22}(u) -l_{12}(qu) l_{21}(u), \quad P^{-}=-\frac{1}{2}R(q^{-1}).
\een
It allows one to define the inverse of the quantum monodromy matrix, in term of the quantum monodromy co-matrix 
\ben\label{hL}
\widehat L(u)=\sigma^{y}L^t(qu)\sigma^{y}=\left(\begin{array}{cc}
      l_{22}(q u) & -l_{12}(qu) \\
  - l_{21}(qu)  & l_{11}(qu) 
      \end{array}
\right),
\een
by the relation $L^{-1}(u)=\frac{\widehat L(uq^{-2})}{{\rm Det}_q\{L(u)\}}$.

The Heisenberg XXZ spin-$\frac{1}{2}$ chain belongs to the fundamental representation of $U_q(\widehat{gl_2})$. 
In this representation the monodromy matrix (\ref{L}) is given by the product of R matrices 
\ben\label{LtoR}
L_a(u)=R_{a1}(u/v_1)\dots R_{aN}(u/v_N)
\een
where $\bar v=\{v_1,\dots v_N\}$ are the so-called inhomogeneity parameters that are crucial in the SoV and ODBA approaches. In this representation the quantum monodromy co-matrix (\ref{hL}) is given by 
\ben\label{hLtoR}
\widehat L_a(q^{-2}u^{-1})=(-1)^NR_{aN}(uv_N)\dots R_{a1}(uv_1).
\een
%
%
\subsection{Coideal sub-algebra of $U_q(\widehat {gl_2})$ in the reflection algebra realization and transfer matrix}
Let us recall the K-matrix \cite{dVGR1}
\ben\label{Km}
K^-(u)=\lt(\begin{array}{cc}k^-(u)&\tau \,c(u)\\
\tilde \tau \,c(u)&k^-(u^{-1})\end{array}\rt),\quad k^-(u) =\nu_-u+\nu_+u^{-1},\quad  c(u)=u^2-u^{-2},
\een 
with parameters $\{\nu_\pm,\tau,\tilde \tau \}\in \CC^4$ related to the right boundary parameters $\{\nu,\tau^\pm\}$ of the Hamiltonian (\ref{H}) (see below (\ref{parKH1})). 
This is the most general solution of the reflection equation
 \ben \label{RE}
 R_{ab}(u_1/u_2)K^-_a(u_1)R_{ab}(u_1u_2)K^-_b(u_2)=
K^-_b(u_2)R_{ab}(u_1u_2)K^-_a(u_1)R_{ab}(u_1/u_2).
\een

The dual K-matrix is given by
\ben\label{Kp}
&&K^+(u)=
\lt(\begin{array}{cc}
k^+(qu)&\tilde \kappa \,c(qu)\\
\kappa \,c(qu)&k^+(q^{-1}u^{-1})
\end{array}\rt),\quad k^+(u) =\epsilon_+u+\epsilon_-u^{-1},
\een
with parameters $\{\epsilon_\pm,\kappa, \tilde \kappa \}\in \CC^4$ that are related to the left boundary parameters $\{\epsilon, \kappa^\pm\}$ of the Hamiltonian (\ref{H}) (see below (\ref{parKH1})).
This is the most general solution of the dual reflection equation 
\ben\label{DRE}
&&R_{ab}(u_2/u_1)K^+_a(u_1)R_{ab}(q^{-2}u^{-1}_1u^{-1}_2)K^+_b(u_2)= K^+_b(u_2)R_{ab}(q^{-2}u^{-1}_1u^{-1}_2)K^+_a(u_1)R_{ab}(u_2/u_1).
\een 

From  the quantum monodromy matrix (\ref{L}), the quantum monodromy co-matrix (\ref{hL}) and the K-matrix (\ref{Km}), one can construct the 
double-row monodromy matrix using the Sklyanin dressing procedure
\footnote{The normalization $(-1)^N{\rm Det}_q\{L(qu^{-1})\}$ is used such that the XXX limit fits with the notations in \cite{BC13}.}  
\ben\label{K}
K_a(u)&=&\big((-1)^N{\rm Det}_q\{L(qu^{-1})\}\big)\, L_a(u)K_a^-(u) \big(L_a(u^{-1})\big)^{-1}\\
&=&\left(\begin{array}{cc}
       \mathscr{A}(u) & \mathscr{B}(u)\\
       \mathscr{C}(u) & \mathscr{D}(u)+\frac{1}{b(qu^2)}\mathscr{A}(u)
      \end{array}
\right)_a,
\een 
where the operators $\{ \mathscr{A}(u) , \mathscr{B}(u), \mathscr{C}(u) , \mathscr{D}(u)\}$ act on the quantum space $\cH$. 
From the dual K-matrix (\ref{Kp}) and the double-row monodromy matrix  (\ref{K}), one can construct the transfer matrix 
\ben\label{tr}
&&t(u)=tr_a(K^+_a(u)K_a(u))=\phi(u)k^+(u)\mathscr{A}(u) +k^+(q^{-1}u^{-1})\mathscr{D}(u)+c(qu)\big(\kappa \,\mathscr{B}(u)+\tilde\kappa \,\mathscr{C}(u)\big)
\een
with
\ben\label{phi}
&& \phi(u)= \frac{b(q^2u^2)}{b(qu^2)}.\nonumber
\een
The transfer matrix commutes for  different spectral parameters \cite{Skly88}, {\it i.e.} $[t(u),t(v)]=0$. 
Thus $t(u)$ is the generating function of the conserved quantities of the model. 
In particular, the Hamiltonian (\ref{H}) can be recovered using the standard formula 
\ben\label{Htotr}
H=\frac{q-q^{-1}}{2} \frac{d}{du}\ln(t(u))\Big|_{u=1, v_i=1}-\Big(N\,\frac{q+q^{-1}}{2}+\frac{(q-q^{-1})^2}{2(q+q^{-1})}\Big).
\een
The relations between the boundary parameters of K-matrices (\ref{Km},\ref{Kp}) and the ones of the Hamiltonian (\ref{H})  are given by
\ben\label{parKH1}
&&\epsilon=\frac{(q-q^{-1})}{2}\frac{(\epsilon_+ - \epsilon_-)}{(\epsilon_+ + \epsilon_-)}
,\quad
  \kappa^-=\frac{2(q-q^{-1})}{(\epsilon_+ + \epsilon_-)}\kappa,\quad   \kappa^+=\frac{2(q-q^{-1})}{(\epsilon_+ + \epsilon_-)}\tilde \kappa,\\
  \label{parKH2}
&& \nu=  \frac{(q-q^{-1})}{2}\frac{(\nu_- - \nu_+)}{(\nu_+ +\nu_-)},\quad
 \tau^-= \frac{2(q-q^{-1})}{(\nu_+ + \nu_-)}\tilde \tau,\quad
 \tau^+= \frac{2(q-q^{-1})}{(\nu_+ + \nu_-)}\tau.
 \een


\subsection{$U_q(\widehat{gl_2})$ highest weight vector and upper triangular right boundary.} 
For finite dimensional representation of the quantum monodromy matrix (\ref{L}) we always have a highest weight representation \cite{Tar85}. 
For the fundamental representation (\ref{LtoR}), the highest weight vector is given by  
\ben\label{Om}
|\Omega\rangle=\otimes_{k=1}^{N}\left(\begin{array}{l}
       1 \\
      0       \end{array}\right)_k \in \cW_0.
     \een
The action of the entries of the quantum monodromy matrix (\ref{L}) on this vector are given by   
\ben\label{lijvac}
&& l_{11}(u)|\Omega\rangle=\lambda_1(u)|\Omega\rangle=\prod_{i=1}^Nb(qu/v_i)|\Omega\rangle, \quad
l_{22}(u)|\Omega\rangle=\lambda_2(u)|\Omega\rangle=\prod_{i=1}^Nb(u/v_i)|\Omega\rangle, \quad
l_{21}(u)|\Omega\rangle=0. 
\een
An important point is that the operator $l_{12}(u)$ is nilpotent on this vector 
\ben\label{l12vac}
&&l_{12}(\bar u)|\Omega\rangle=l_{12}(u_1)\dots l_{12}(u_N)|\Omega\rangle=Z(\bar u,\bar v)|\hat  \Omega\rangle \quad \mbox{and} \quad l_{12}(u) |\hat  \Omega\rangle=0
\een
with
\ben\label{bOm}
|\hat \Omega\rangle=\otimes_{k=1}^{N}\left(\begin{array}{l}
       0 \\
      1       \end{array}\right)_k \in \cW_N
\een
the lowest weight vector, and
\ben\label{Z}
Z(\bar u|\bar v)=\frac{ {\rm Det}\{a(u_i,v_j)\}}{a(\bar u,\bar v)\prod_{i<j}b(u_i/u_j)b(v_j/v_i) } ,\quad a(u_i,v_j)=\frac{1}{b(u_i/v_j)b(qu_i/v_j)},
\een
the domain wall partition function of the trigonometric six vertex model given in term of the Izergin determinant \cite{Ize87}.

For the case that the right boundary is upper triangular  $\tilde \tau=0$, we can use  the definition of the quantum double row monodromy matrix (\ref{K}) to find the expression of the operators  $\{\mathscr{A}(u),\mathscr{B}(u),\mathscr{C}(u),\mathscr{D}(u)\}$ in term of the operators  $\{l_{ij}(u)\}$, namely
\ben\label{Atolij}
&&\mathscr{A}(u)=(-1)^N\big(k^-(u)l_{11}(u)l_{22}(q^{-1}u^{-1})-k^-(u^{-1})l_{12}(u)l_{21}(q^{-1}u^{-1})\big)\\
&&\qquad\qquad\qquad\qquad+(-1)^N\tau c(u)\big(\frac{1}{b(q u^2)}l_{21}(u)l_{11}(q^{-1}u^{-1})-\phi(u) l_{21}(q^{-1}u^{-1})l_{11}(u)\big),\nonumber\\ \label{Btolij}
&&\mathscr{B}(u)=(-1)^N\phi(q^{-1}u^{-1})\Big(k^-(q^{-1}u^{-1})l_{12}(u)l_{11}(q^{-1}u^{-1})-k^-(u)l_{12}(q^{-1}u^{-1})l_{11}(u)\Big)\\ 
&&\qquad\qquad\qquad\qquad+(-1)^N\tau c(u)l_{11}(u)l_{11}(q^{-1}u^{-1}),\nonumber\\
&&\mathscr{C}(u)= (-1)^N\phi(q^{-1}u^{-1})\Big(k^-(qu)l_{21}(u)l_{11}(q^{-1}u^{-1})-k^-(u^{-1})l_{21}(q^{-1}u^{-1})l_{11}(u)\Big) \label{Ctolij}\\ 
&&\qquad\qquad\qquad\qquad-(-1)^N\tau c(u)l_{21}(u)l_{21}(q^{-1}u^{-1}),\nonumber\\
&&\mathscr{D}(u)=(-1)^N\phi(q^{-1}u^{-1})\big(k^-(q^{-1}u^{-1})l_{11}(q^{-1}u^{-1})l_{22}(u)-k^-(qu)l_{12}(q^{-1}u^{-1})l_{21}(u)\big)\label{Dtolij}\\
&&\qquad+(-1)^N\tau c(q^{-1}u^{-1})\phi(q^{-1}u^{-1})\big(\frac{1}{b(q^{-1} u^{-2})}l_{21}(q^{-1}u^{-1})l_{11}(u)-\phi(q^{-1}u^{-1}) l_{21}(u)l_{11}(q^{-1}u^{-1})\big).\nonumber
\een
It follows from (\ref{lijvac}) the actions on the highest weight vector (\ref{Om})
\begin{eqnarray}\label{Avac}
 &&\mathscr{A}(u)|\Omega\rangle=k^-(u) \Lambda(u)|\Omega\rangle,\quad
  \mathscr{D}(u)|\Omega\rangle=\phi(q^{-1}u^{-1})k^-(q^{-1}u^{-1})\Lambda(q^{-1}u^{-1})|\Omega\rangle, \quad \mathscr{C}(u)|\Omega\rangle=0,
\end{eqnarray}
with
\ben
 \Lambda(u) =(-1)^N\lambda_1(u)\lambda_2(q^{-1}u^{-1})= \prod_{j=1}^{N}b(qu/v_i)b(quv_i).
\een
\begin{rmk} \label{ZD} 
 For $\tau=0$ (i.e. for diagonal $K^-$ matrix) one has a nilpotent property for the $\mathscr{B}(u)$ \cite{Tsu98}
\ben
\mathscr{B}(u_1)\dots \mathscr{B}(u_N)|\Omega\rangle= Z_d(\bar u| \bar v)|\hat  \Omega\rangle \quad \mbox{and} \quad  \mathscr{B}(u) |\hat  \Omega\rangle=0
\een
with
\ben
&&Z_d(\bar u| \bar v)=(-1)^N\frac{\prod_{i,j=1}^Nb(u_i/v_j)b(u_iv_j)b(qu_i/v_j)b(qu_iv_j)}{\prod_{i<j}b(u_i/u_j)b(qu_iu_j)b(v_j/v_i)b(v_iv_j)} {\rm Det}\Big(M(u_i,v_j)\Big)\\
&&M(u,v)=\frac{\phi(q^{-1}u^{-1})}{b(q u v) b(v/u)}\Big(\frac{k^-(u)}{b(uv)}+\frac{k^-(q^{-1}u^{-1})}{b(q u/v)}\Big)
\een
the partition function of the trigonometric six vertex model with domain wall boundary conditions and one diagonal reflecting end. Moreover in this case the vector
\ben
\Phi_{d}^M(\bar u)=\mathscr{B}(\bar u)|\Omega\rangle=\mathscr{B}(u_1)\dots\mathscr{B}(u_M)|\Omega\rangle
\een
is an element of the subspace $\cW_M$. If we consider the same vector with $\tau\neq0$, it belongs to $\cW_M\oplus\cW_{M-1}\oplus \dots \oplus\cW_0$. Thus for $M=N$ it belongs to the full quantum space $\cH$.
\end{rmk}
\begin{rmk} \label{FVtf}  For a general $K^-$ matrix, a Face-Vertex transformation \cite{baxter} can be performed such that a highest weight vector can be found for the new operators that belong into a dynamical version of the Reflection algebra, see \cite{gauge,gaugeKF} and references therein. 
For some specific Face-Vertex transformation the nilpotent property for the dynamical creation operator can be obtained together with the partition function of the trigonometric six vertex model with domain wall boundary conditions and a non-diagonal reflecting end, see \cite{gaugeKF}. 
Such type of transformation will be considered to perform the MABA for the Heisenberg XXZ spin-$\frac{1}{2}$ chain on the segment with generic boundaries \cite{BP2}.
\end{rmk}
%


\section{Algebraic Bethe ansatz : diagonal/diagonal and diagonal/upper triangular cases \label{S:UABA}}
For these cases that the left boundary is diagonal $\kappa=\tilde \kappa=0$ and the right boundary is diagonal $\tau=\tilde \tau=0$ or upper triangular  $\tilde \tau=0$, we say that the transfer matrix (\ref{tr}) is diagonal
\footnote{{\it i.e.} it only involves diagonal elements $\mathscr{A}(u)$ and $\mathscr{D}(u)$ of the double row quantum monodromy matrix (\ref{K}).}
and is given by
\ben\label{td}
t(u)=t_d(u)=\phi(u)k^+(u)\mathscr{A}(u) +k^+(q^{-1}u^{-1})\mathscr{D}(u).
\een
Since the action on the highest weight vector of  $\mathscr{A}(u)$ and $\mathscr{D}(u)$ does not depend on $\tau$, the generalized ABA introduced by Sklyanin \cite{Skly88} can be also applied to diagonal/upper triangular case \cite{MRM}. 
The Bethe Vectors are given by
\ben\label{BVd}
\Phi_{d}^M(\bar u)=\mathscr{B}(\bar u)|\Omega\rangle=\mathscr{B}(u_1)\dots\mathscr{B}(u_M)|\Omega\rangle.
\een
with $M\in\{0,1,\dots,N\}$. 
From the commutation relations (\ref{comAB},\ref{comDB},\ref{comBB}) and the actions on the highest weight vector (\ref{Avac}), one can show the actions
\ben\label{acAB}
&&\qquad\qquad\qquad\mathscr{A}(u)\Phi_{d}^M(\bar u)=k^-(u)\Lambda(u)f(u,\bar u)\Phi_{d}^M(\bar u)\\
&&+\sum_{j=1}^M\Big(g(u,u_j)k^-(u_j)\Lambda(u_j)f(u_j,\bar u_j)+w(u,u_j)\phi(q^{-1}u_j^{-1})k^-(q^{-1}u_j^{-1})\Lambda(q^{-1}u_j^{-1})h(u_j,\bar u_j)\Big)\Phi_{d}^M(\{u,\bar u_i\}),\nonumber
\een
and
\ben\label{acDB}
&&\qquad\qquad\qquad\mathscr{D}(u)\Phi_{d}^M(\bar u)=\phi(q^{-1}u^{-1})k^-(q^{-1}u^{-1})\Lambda(q^{-1}u^{-1})h(u,\bar u)\Phi_{d}^M(\bar u)\\
&&+\sum_{j=1}^N\big(l(u,u_j)\phi(q^{-1}u_j^{-1})k^-(q^{-1}u_j^{-1})\Lambda(q^{-1}u_j^{-1})h(u_j,\bar u_j)+n(u,u_j)k^-(u_j)\Lambda(u_j)f(u_j,\bar u_j)\Big)\Phi_{d}^M(\{u,\bar u_i\}).\nonumber
\een
It follows, using the relation (\ref{idUWT1},\ref{idUWT2}), the off-shell equation\footnote{{\it i.e.} the  parameters $\bar u$, with $\# \bar u=M$, are arbitrary.} for the action of the diagonal transfer matrix (\ref{td})%
\ben\label{Off-d}
&&t_d(u)\Phi_{d}^M(\bar u)=
\Lambda_d^M(u,\bar u)\Phi_{d}^M(\bar u)+ \sum_{i=1}^{M}F(u, u_i)E^M_d(u_i,\bar u_i)\Phi_{d}^{M}(\{u,\bar u_i\})
\een
with
\ben
&& \Lambda^M_d(u,\bar u)=\psi(u)f(u,\bar u)+\psi(q^{-1}u^{-1})h(u,\bar u), \label{Ed}
\\
&&E^M_d(u_i,\bar u_i)=\phi(q^{-1}u_i^{-1})\psi(u_i)f(u_i,\bar u_i)-\phi(u_i)\psi(q^{-1}u_i^{-1})h(u_i,\bar u_i)=\lim_{u \to u_i}\Big(b(u_i/u)\Lambda^M_d(u,\bar u)\Big),
\een
and
\ben
\psi(u)=\phi(u)k^+(u)k^-(u)\Lambda(u).
\een
The eigenvectors of the transfer matrix follow by imposing that the arbitrary parameters $\bar u$ satisfy the Bethe equations $E^M_d(u_i,\bar u_i)=0$ with $i=1,\dots M$. In this case the BV, $\Phi_{d}^M(\bar u)$, is said on-shell.  

\begin{rmk} \label{SCBB}
For $\tau\neq 0$ this is an example of model that does not have $U(1)$ symmetry but where usual ABA can be applied. In this case the action of the operator $\mathscr{B}(u)$ on the BV with $M=N$ will have a non-trivial off-shell structure that will be given in section \ref{S:MABA}. This will allow one to study correlation functions of the form 
\ben
S_{up}^{P+M}(\bar w |\bar u) =\langle \hat \Omega|\mathscr{B}(\bar w)\mathscr{B}(\bar u)| \Omega\rangle
\een
with $\# \bar w=P$ and $\# \bar u=M$. For $P+M=N$ we have $S_{up}^{N}(\bar w |\bar u) = Z_d(\{\bar w ,\bar u\}|\bar v)$ and for $P+M<N$ we have $S_{up}^{P+M}(\bar w |\bar u) =0$  .  
\end{rmk}




\section{Toward the Modified algebraic Bethe ansatz : upper/upper triangular case \label{S:MABAupup}}
For this case that left boundary is upper triangular $\kappa=0$ and that right boundary is also upper triangular $\tilde \tau=0$, the transfer matrix has an off-diagonal term that involves the operator $\mathscr{C}(u)$ and is given by
\ben\label{tup}
t_{up}(u)=t_d(u)+\tilde \kappa  c(qu)\mathscr{C}(u). 
\een
The highest weight vector (\ref{Om}) is a highest weight vector as for the diagonal transfer matrix (\ref{tup}). 
The Bethe vectors in on-shell case have been first derived in \cite{pimenta} extending the result for the XXX chain given in \cite{BCR12}. 
Here we give a factorized formula for the Bethe Vectors and a derivation of the result based on the usual ABA technique but with new operators that depend of an integer $m$, namely 
\ben
&&\widetilde{\mathscr{A}\null}(u,m)=\mathscr{A}(u)-q^{m}u^{-1}\frac{\tilde \kappa}{q \epsilon_-}\,\mathscr{C}(u), \label{Atfuu}\\
&&{\widetilde{\mathscr{D}}}(u,m) =\mathscr{D}(u)+q^{m}\,q u\,\frac{\tilde \kappa}{q \epsilon_-}\phi(u)\mathscr{C}(u), \label{Dtfuu}\\
&&{\widetilde{\mathscr{B}}}(u,m)=\mathscr{B}(u)+q^{m+2}\frac{\tilde \kappa}{q \epsilon_-}\Big(q u\frac{b(u^2)}{b(qu^2)}\mathscr{A}(u)-u^{-1}\,\mathscr{D}(u)\Big)-\Big(q^{m+2}\frac{\tilde \kappa}{q\epsilon_-}\Big)^2\mathscr{C}(u) \label{Btfuu}.
\een
The actions of $\widetilde{\mathscr{A}\null}(u,m)$ and ${{\widetilde{\mathscr{D}}}}(u,m)$  on the highest weight vector are the same than for ${\mathscr{A}}(u)$ and ${{\mathscr{D}}}(u)$
\ben
&&\widetilde{\mathscr{A}\null}(u,m)|\Omega\rangle=k^-(u) \Lambda(u)|\Omega\rangle,\\
&&{{\widetilde{\mathscr{D}}}}(u,m)|\Omega\rangle=\phi(q^{-1}u^{-1})k^-(q^{-1}u^{-1})\Lambda(q^{-1}u^{-1})|\Omega\rangle.
\een
From the commutation relations given in the appendix \ref{App:Func} we can show the commutation relations 
\ben 
&&{\widetilde{\mathscr{B}}}(u,m){\widetilde{\mathscr{B}}}(v,m-2)={\widetilde{\mathscr{B}}}(v,m){\widetilde{\mathscr{B}}}(u,m-2), \label{BBtf}\\  
&&\widetilde{\mathscr{A}\null}(u,m+2){\widetilde{\mathscr{B}}}(v,m)=f(u,v){\widetilde{\mathscr{B}}}(v,m)\widetilde{\mathscr{A}\null}(u,m)+g(u,v){\widetilde{\mathscr{B}}}(u,m)\widetilde{\mathscr{A}\null}(v,m)+w(u,v){\widetilde{\mathscr{B}}}(u,m){{\widetilde{\mathscr{D}}}}(v,m),\label{ABtf}\\  
&&{{\widetilde{\mathscr{D}}}}(u,m+2){\widetilde{\mathscr{B}}}(v,m)=h(u,v){\widetilde{\mathscr{B}}}(v,m){{\widetilde{\mathscr{D}}}}(u,m)+k(u,v){\widetilde{\mathscr{B}}}(u,m){{\widetilde{\mathscr{D}}}}(v,m)+n(u,v){\widetilde{\mathscr{B}}}(u,m)\widetilde{\mathscr{A}\null}(v,m).\label{DBtf}
\een 
These new operators are related to the Face-Vertex transformation mentioned in the remark \ref{FVtf}, more details on this point will be given in \cite{BP2}. 
The transfer matrix (\ref{tup}) can be rewritten in a modified diagonal form using these new operators, namely
\ben \label{tupm}
t_{up}(u)=\phi(u)k^+(u)\widetilde{\mathscr{A}\null}(u,0) +k^+(q^{-1}u^{-1}){\widetilde{\mathscr{D}}}(u,0). 
 \label{tuptf}
\een
Then, we can introduce the BV 
\ben
&&\Phi_{up}^M(\bar u)={\widetilde{\mathscr{B}}}(u_1,-2){\widetilde{\mathscr{B}}}(u_2,-4)\dots{\widetilde{\mathscr{B}}}(u_M,-2M)|\Omega\rangle \label{BVup}
\een
with $M\in\{0,1,\dots,N\}$ and that are,  from (\ref{BBtf}), symmetric functions of the parameters $\bar u$.  We can show that actions (\ref{acAB}) and (\ref{acDB}) with $\Phi_{d}^M(\bar u)\to \Phi_{up}^M(\bar u)$, ${\mathscr{A}}(u)\to \widetilde{\mathscr{A}\null}(u,0)$ and ${\mathscr{D}}(u)\to {{\widetilde{\mathscr{D}}}}(u,0)$ are valid. Thus, the same steps as in the previous section
permit to obtain the off-shell equation for the modified transfer matrix (\ref{tupm})
\ben
t_{up}(u)\Phi_{up}^M(\bar u)=\Lambda_d^M(u,\bar u)\Phi_{up}^M(\bar u)+\sum_{i=1}^MF(u, u_i)E^M_d(u_i,\bar u_i)\Phi_{up}^M(u,\bar u_i). \label{Offup}
\een
The eigenvectors of the transfer matrix follow by imposing that the arbitrary parameters $\bar u$ satisfy the Bethe equations $E^M_d(u_i,\bar u_i)=0$ with $i=1,\dots M$. In this case the BV, $\Phi_{up}^M(\bar u)$, is said on-shell.  
\begin{rmk}
The BV (\ref{BVup})  are linear combinations of the ones of the previous section. 
Using (\ref{Btfuu}), the commutation relations in the appendix \ref{App:Func} and the action  on the highest weight vector (\ref{Avac}), one will find   
\ben
\Phi_{up}^M(\bar u)=\sum_{i=0}^M \sum_{\bar u\Rightarrow\{\bar u_{\so},\;\bar u_{\st}\}} W_i(\bar u_{\so}|\bar u_{\st}) \Phi_{d}^i(\bar u_{\st}).
\een  
The second sum corresponds to each splitting of the set $\bar u$ into subsets $\bar u_{\so}$ and $\bar u_{\st}$ with $\#\bar u_{\st}=i$ and the elements in every subset are ordered in such a way that the sequence of their subscripts is strictly increasing. 
The explicit form of the $W_i(\bar u_{\so}|\bar u_{\st})$ is given in \cite{pimenta} for the on-shell case. 
\end{rmk}

\begin{rmk}
In the XXX limit, this result gives a factorized form and a simplest proof for the Bethe vectors given in \cite{BCR12} for the on-shell case and in \cite{CMS14} for the off-shell case where recursion relation for the $W_i(\bar u_{\so}|\bar u_{\st})$ are given.  
In this limit the new operators are independent of $m$, namely 
\ben
&&\widetilde{\mathscr{A}\null}(\lambda)=\mathscr{A}(\lambda)+\frac{ \xi^+}{2}\,\mathscr{C}(\lambda),\quad {\widetilde{\mathscr{D}}}(\lambda) =\mathscr{D}(\lambda)-\frac{\xi^+}{2}\frac{2(\lambda+1)}{2\lambda+1}\mathscr{C}(\lambda), \\
&&{\widetilde{\mathscr{B}}}(\lambda)=\mathscr{B}(\lambda)-\frac{\xi^+}{ 2}\Big(\frac{2\lambda}{2\lambda+1}\mathscr{A}(\lambda)-\mathscr{D}(\lambda)\Big)-\Big(\frac{\xi^+}{ 2}\Big)^2\mathscr{C}(\lambda).
\een
Thus the BV are given by
\ben
&&\Phi_{up}^M(\bar \lambda)={\widetilde{\mathscr{B}}}(\lambda_1){\widetilde{\mathscr{B}}}(\lambda_2)\dots{\widetilde{\mathscr{B}}}(\lambda_M)|\Omega\rangle. 
\een
This result was already pointed out in \cite{BC13} from the limit of the generic boundary BV.
\end{rmk}



\section{Modified algebraic Bethe ansatz : lower/upper triangular case  \label{S:MABA}}

In this case that left boundary is lower triangular $\tilde \kappa=0$ and that right boundary is upper triangular $\tilde \tau=0$, the transfer matrix has an off-diagonal term that involves the $\mathscr{B}(u)$ operator and is given by
\ben\label{trlp}
t_{lo/up}(u)=t_d(u)+\kappa c(qu)\mathscr{B}(u). 
\een
The highest weight vector (\ref{Om}) is not an eigenvector of this transfer matrix. Indeed, acting with the transfer matrix (\ref{tr}) one finds
\ben
t_{lo/up}(u)\Om=\Big(\psi(u)+\psi(q^{-1}u^{-1})\Big)\Om+\kappa\, c(qu) \mathscr{B}(u)|\Omega\rangle,
\een
that have a term with a $\mathscr{B}(u)$ operator. It will be the same for all vectors of the form (\ref{BVd}) with $M\neq N$. 
Let us consider the Bethe vector
\ben\label{BVlo/up}
\Phi^N_{lo/up}(\bar u)=\mathscr{B}(\bar u)|\Omega\rangle
\een
with $\#\bar u=N$. The action of diagonal part of the transfer matrix (\ref{trlp}) on this vector is given by (\ref{Off-d}) with $M=N$. 
The number of creation operators corresponds to the length $N$ of the chain, thus as it was said in remark \ref{ZD} this vector belongs to the full quantum space 
\ben
\Phi^N_{lo/up}(\bar u)\in\cH=\oplus_{i=0}^{N} \cW_i.
\een 
The new hypothesis that leads to the modified version of the ABA is that the action of the creation operator $\mathscr{B}(u)$ on this vector has an off-shell action of the form
\ben\label{BonBV}
\kappa \,c(qu) \mathscr{B}(u)\Phi_{lo/up}^N(\bar u)=\Lambda_g^N(u,\bar u)\Phi_{lo/up}^N(\bar u)+\sum_{i=1}^NF(u,u_i)E_g^N(u_i,\bar u_i)\Phi_{lo/up}^N(\{u,\bar u_i\}),
\een
similar to the one of the diagonal part (\ref{Off-d}).
Considering small $N=1,2$ cases given in the next section, it shows that such action exists and allows one to conjecture for generic $N$ that
\ben
&& \Lambda_g^N(u,\bar u)=-\tau\,\kappa\, c(u)c(q^{-1}u^{-1}) \Lambda(u)\Lambda(q^{-1}u^{-1})m(u,\bar u), \label{Eg}\\
&& E_g^N(u_i,\bar u_i)=\tau\,\kappa\, \frac{c(u_i)c(q^{-1}u_i^{-1})}{b(q u_i^2)} \Lambda(u_i)\Lambda(q^{-1}u_i^{-1})m(u_i,\bar u_i) =\lim_{u\to u_i}\Big(b(u_i/u)\Lambda^N_g(u,\bar u)\Big).
\een
\begin{rmk}
For $\tau=0$, one recovers the nilpotent property given in remark \ref{ZD}.
\end{rmk}
\begin{rmk}\label{SCBB2}
We can use this off-shell action for the creation operator $\mathscr{B}(u)$ to calculate the recursion relation for the scalar product given in remark \ref{SCBB}
\ben
S_{up}^{P+M}(\bar w |\bar u) = \Lambda_g^N(w_j,\bar u)S_{up}^{P+M-1}(\bar w_j |\bar u)+\sum_{i=1}^NF(w_j,u_i) E_g^N(u_i,\bar u_i)S_{up}^{P+M-1}(\bar w |\bar u_i).
\een
with $\# \bar w=P$, $\# \bar u=M$ and $P+M>N$. 
\end{rmk}
Finally from  (\ref{Off-d}) and (\ref{BonBV}), we arrived to the main result of the paper
\ben\label{tronBV}
&&t_{lo/up}(u)\Phi^N_{lo/up}(\bar u)=\Lambda^N(u,\bar u)\Phi^N_{lo/up}(\bar u)+\sum_{i=1}^NF(u,u_i)E^N(u_i,\bar u_i)\Phi^N_{lo/up}(\{u,\bar u_i\})
\een
with
\ben
&&\Lambda^N(u,\bar u)=\Lambda^N_d(u,\bar u)+\Lambda^N_g(u,\bar u),\quad E^N(u_i,\bar u_i) =E_d^N(u_i,\bar u_i)+E_g^N(u_i,\bar u_i).\label{LamE} 
\een
The eigenvectors of the transfer matrix follow by imposing that the arbitrary parameters $\bar u$ satisfy the Bethe equations $E^N(u_i,\bar u_i)=0$.  $\Lambda^N(u,\bar u)$ has an additional term and satisfy all the relations used in the ODBA \cite{CYSW3}. 
\begin{rmk}
In the XXX limit this result gives, up to a similarity transformation, the solution for general left and right boundaries and provides an alternative presentation for the Bethe vectors found in \cite{BC13}. Moreover it gives the conjecture independently of the knowledge of the eigenvalues that has been used in \cite{BC13}.  Let us also mention that we realized, {\it  a posteriori}, that all ingredients to obtain the conjecture independently of the eigenvalues were already present in \cite{BC13}, the key step was the introduction of the new operators that gives the modified diagonal transfer matrix. This will be presented in \cite{BP}.   
\end{rmk}
\begin{rmk}\label{Noplu}
As in the previous section, we can introduce new operators to put the transfer matrix in a modified diagonal form 
\ben\label{trlpd}
t_{lo/up}(u)&=&\phi(u)k^+(u){\overline{\mathscr{A}}}(u,0) +k^+(q^{-1}u^{-1})\overline{\mathscr{D}}(u,0) 
\een
with
\ben\label{ADtf}
&&{\overline{\mathscr{A}}}(u,m)= \mathscr{A}(u)-q^{m}\frac{\kappa}{q\epsilon_-} u^{-1}\mathscr{B}(u),\quad \overline{\mathscr{D}}(u,m) = \mathscr{D}(u)+q^{m}\frac{\kappa}{q\epsilon_-}{q u}\phi(u)  \mathscr{B}(u).
\een
These new operators have off-diagonal actions on the highest weight vector 
\ben\label{AtfonVac}
&&{\overline{\mathscr{A}}}(u,m)|\Omega\rangle= k^-(u) \Lambda(u)|\Omega\rangle-q^{m} u^{-1}\frac{\kappa}{q\epsilon_-}\mathscr{B}(u)|\Omega\rangle,\\ \label{DtfonVac}
&&\overline{\mathscr{D}}(u,m)|\Omega\rangle= \phi(q^{-1}u^{-1})k^-(q^{-1}u^{-1})\Lambda(q^{-1}u^{-1})|\Omega\rangle+q^{m}{q u}\frac{\kappa}{q\epsilon_-}\phi(u) \mathscr{B}(u)|\Omega\rangle.
\een
Such type of off-diagonal action was already pointed out in XXX case where new operators were also introduced \cite{BC13}.
These new operators have commutation relations 
\ben 
&&{\overline{\mathscr{A}}}(u,m+2){\mathscr{B}}(v)=f(u,v){\mathscr{B}}(v){\overline{\mathscr{A}}}(u,m)+g(u,v){\mathscr{B}}(u){\overline{\mathscr{A}}}(u,m)+w(u,v){\mathscr{B}}(u){\overline{\mathscr{D}}}(v,m),\label{ABtflu}\\  
&&{\overline{\mathscr{D}}}(u,m+2){\mathscr{B}}(v)=h(u,v){{\mathscr{B}}}(v){\overline{\mathscr{D}}}(u,m)+k(u,v){\mathscr{B}}(u){\overline{\mathscr{D}}}(v,m)+n(u,v){\mathscr{B}}(u){\overline{\mathscr{A}}}(u,m).\label{DBtflu}
\een 
Acting with the transfer matrix (\ref{trlpd}) on Bethe Vector (\ref{BVlo/up}), we can show from  commutation relations (\ref{ABtflu},\ref{DBtflu}) and actions on the highest weight vector (\ref{AtfonVac},\ref{DtfonVac}) that
\ben\label{tronBVbis}
&&t_{lo/up}(u)\Phi^N_{lo/up}(\bar u)=\Lambda_d^N(u,\bar u)\Phi^N_{lo/up}(\bar u)+\sum_{i=1}^NF(u,u_i)E_d^N(u_i,\bar u_i)\Phi^N_{lo/up}(\{u,\bar u_i\})\\
&&\qquad \qquad\qquad \qquad\qquad +\kappa \,c(qu) \mathscr{B}(u)\Phi_{lo/up}^N(\bar u).\nonumber
\een
Then, using the conjecture (\ref{BonBV}) of the action of the $ \mathscr{B}(u)$ operator on the Bethe vector (\ref{BVlo/up}) we arrive to (\ref{tronBV}). This way of considering the problem is quite artificial here but will be crucial for the general boundaries case \cite{BP2}.
\end{rmk}



\section{Construction of the conjecture from small cases \label{S:smallN}}
In this section we use the notation
\ben
\tilde l_{12}(u)=k^-(q^{-1}u^{-1})\lambda_{1}(q^{-1}u^{-1})l_{12}(u).
\een
To construct the conjecture we use the explicit form of the operator $\mathscr{B}(u)$ in term of $l_{ij}(u)$ operators (\ref{Btolij}). 
Then imposing the off-shell action to be of the form (\ref{BonBV}) we order the operators using commutation relations given in the appendix \ref{App:Func} and project on the basis 
\ben \label{basis} 
\{ \Om,  \quad l_{12}(u_i) \Om,  \quad  l_{12}(u_i)  l_{12}(u_j)\Om,   \quad  l_{12}(u_i)  l_{12}(u_j)  l_{12}(u_k)\Om, \quad \dots, \quad \bar \Om\}.
\een
For a set of formal parameters $\bar u=\{u_1,\dots,u_N\}$, with $u_i \neq u_j\neq u_k \neq\dots$, this basis has dimension $2^N$ and provide a basis for $\cH$. 
From this procedure we obtain a set of equations that allows one to fix $\Lambda^N_g(u,\bar u)$ and $E^N_g(u_i,\bar u_i)$ that we consider as independent unknowns. 
\begin{rmk}\label{proj}
For general $N$, we have to project the elements $l_{12}(w)l_{12}(u_{j_2})\dots l_{12}(u_{j_{m}})|\Omega \rangle$ with $w \notin \{\bar u\}$ and $0\leq m\leq N$ on the basis (\ref{basis}). 
For a fixed $m$ and $1\leq j_2< \dots <j_m\leq N$, as the basis (\ref{basis}) is complete, we have
\ben
l_{12}(w)l_{12}(u_{j_2})\dots l_{12}(u_{j_{m}})|\Omega\rangle=\sum_{1\leq i_1 < i_2< \dots <i_m\leq N} 
V^{i_1,i_2,\dots,i_m}_{j_2,\dots,j_m}(w|\bar u)\,l_{12}(u_{i_1})\dots l_{12}(u_{i_m})|\Omega\rangle.
\een
For $N=1,2$ the coefficients $V^{i_1,i_2,\dots,i_m}_{j_2,\dots,j_m}(w|\bar u)$ are simple and can be explicitly calculated. 
For generic $N$, only the case $m=N-1$, related to the partition function (\ref{Z}), and the case $m=1$, that corresponds to a Lagrange interpolation of the $l_{12}(u)$ operator at the  points $\bar u$, are simple to calculate. The other coefficients still to be determined 
to prove the conjecture of the off-shell action (\ref{BonBV}) for general $N$ in the way we use for the case $N=1,2$.
However, the case $N=1,2$ are enough to make the conjecture and then one can check numerically $N=3,4$ the off-shell action (\ref{BonBV}) using explicit matrix form of the $\mathscr{B}$ operator to support the conjecture. 
\end{rmk}  
  
\subsection{Case $N=1$}
The BV is given by
\ben\label{BVN1}
\mathscr{B}(u_1)|\Omega\rangle&=&\Big(\phi(q^{-1}u_1^{-1})\big(\tilde l_{12}(q^{-1}u_1^{-1})-\tilde  l_{12}(u_1)\big)-\tau c(u_1)\lambda_{1}(u)\lambda_{1}(q^{-1}u_1^{-1})\Big)|\Omega\rangle,
\een
the nilpotent property by (\ref{l12vac})
\ben
l_{12}(u)l_{12}(v)|\Omega\rangle=0
\een
and we have the partition function (\ref{Z})
\ben
l_{12}(u)|\Omega\rangle=Z(u_1|v_1)|\bar \Omega\rangle=|\bar \Omega\rangle.
\een 
With these actions and the commutation relations (\ref{coml12l12},\ref{coml11l12},\ref{coml22l12},\ref{coml21l12}) we can project the off-shell action (\ref{BonBV}) on the basis (\ref{basis})
\ben
 \{\Om, \quad \bar \Om\}
\een
and obtain two equations for the two unknowns  $\Lambda^1_g(u,u_1)$ and $E^1_g(u_1,\emptyset)$. They give (\ref{Eg}) for $N=1$. 

\subsection{Case $N=2$}
The BV is given by
\ben\label{BVN2}
&&\quad \mathscr{B}(u_1)\mathscr{B}(u_2)|\Omega\rangle=c(u_1)c(u_2)\Big\{\tau^2\lambda_1(u_1)\lambda_1(q^{-1}u_1^{-1})\lambda_1(u_2)\lambda_1(q^{-1}u_2^{-1})\\
&&\qquad \qquad +\tau \frac{ \lambda_1(u_2)\lambda_1(q^{-1}u_2^{-1})}{c(q^{1/2}u_1)}\big(h(u_1,u_2)\tilde l_{12}(u_1) - f(u_1,u_2)\tilde l_{12}(q^{-1}u_1^{-1})\big)\nonumber\\
&&\qquad \qquad +\tau \frac{ \lambda_1(u_1)\lambda_1(q^{-1}u_1^{-1})}{c(q^{1/2}u_2)}\big(h(u_2,u_1)\tilde l_{12}(u_2) - f(u_2,u_1)\tilde l_{12}(q^{-1}u_2^{-1})\big)\Big\}|\Omega\rangle\nonumber\\ 
&&\qquad+\phi(q^{-1}u^{-1}_1)\phi(q^{-1}u^{-1}_2)\Big\{\frac{b(q^2u_1 u_2)}{b(qu_1 u_2)}\tilde l_{12}(u_1)\tilde l_{12}(u_2)+\frac{b(u_1 u_2)}{b(qu_1 u_2)}
\tilde l_{12}(q^{-1}u_1^{-1})\tilde l_{12}(q^{-1}u_2^{-1})\nonumber\\
&&\qquad\qquad-\Big(\frac{b(qu_1 /u_2)}{b(u_1/ u_2)}
\tilde l_{12}(u_1)\tilde l_{12}(q^{-1}u_2^{-1})+\frac{b(qu_2 /u_1)}{b(u_2/ u_1)}
\tilde l_{12}(u_2)\tilde l_{12}(q^{-1}u_1^{-1})\Big)\Big\}|\Omega\rangle,\nonumber
\een
the nilpotent property (\ref{l12vac})  by 
\ben
l_{12}(u)l_{12}(u_1)l_{12}(u_2)|\Omega\rangle=0
\een
and we have the partition function (\ref{Z})
\ben \label{l12vacN2}
l_{12}(u_1)l_{12}(u_2)|\Omega\rangle=Z(u_1,u_2|v_1,v_2) |\bar \Omega\rangle.
\een
To project the off-shell action (\ref{BonBV}) on the basis
\ben
 \{\Om,  \quad l_{12}(u_1) \Om,  \quad l_{12}(u_2) \Om, \quad \bar \Om\}
 \een
we also use the relation
\ben \label{l12N2rela}
l_{12}(u)=\frac{b(u/u_2)}{b(u_1/u_2)}l_{12}(u_1)+\frac{b(u/u_1)}{b(u_2/u_1)}l_{12}(u_2),
\een
that follows from the explicit matrix formulation of $l_{12}(u)$. 
To find $\Lambda^2_g(u,u_1,u_2)$,  $E^2_g(u_1, u_2)$ and $E^2_g(u_2, u_1)$ it is enough to consider the equations from projection on $\Om$ and  $ l_{12}(u_1) \Om$. The first equation gives $\Lambda^2_g(u,u_1,u_2)$ in term of $E^2_g(u_1, u_2)$ and $E^2_g(u_2, u_1)$. Since $E^2_g(u_1, u_2)$ and $E^2_g(u_2, u_1)$ are independent of $u$, we can consider the second equation as a Laurent polynomial in u with each coefficients equal to zero. It provides an overdetermined system of equations for $E^2_g(u_1, u_2)$ and $E^2_g(u_2, u_1)$ that can be solved. It gives (\ref{Eg}) for $N=2$ that allows one to conjecture the case for arbitrary $N$. 

The case $N=3$ has been explicitly checked.   

\section{Conclusion} \label{S:Conc}

We have constructed the BV, eigenvalues and BE for the Heisenberg  XXZ spin-$\frac{1}{2}$ chain on the segment with two upper  and lower/upper triangular boundaries. 
For the former, a factorized formula of the BV and an algebraic proof similar to the usual ABA for the off-shell action of the transfer matrix have been given. 
It relies on the introduction of new operators, linear combination of the ones of the double row quantum monodromy matrix, that allows one to put the transfer matrix in a modified diagonal form.  
For the latter, we have presented a constructive version of the MABA that allows one to fix the BV, eigenvalue and BE. In particular the additional term in the eigenvalues and the BE appears to correspond to the off-shell action of the creation operator on the BV. 
This action was conjectured.
Let us remark that similar results can be obtained for two lower and upper/lower triangular boundaries starting from the lower highest weight vector of $U_q(\widehat{gl_2})$ (\ref{bOm}) to construct the BV.
These results are a first step toward the use of the MABA to conjecture the BV for general/diagonal (or triangular) and general/general boundaries that will be presented in a separate paper \cite{BP2}. This will involve on the one hand the introduction of new operators in the spirit of the ones used for the two upper triangular boundaries (\ref{Atfuu},\ref{Dtfuu},\ref{Btfuu}) and the ones in remark \ref{Noplu} and on the other hand the construction of the off-shell action of the new creation operator on the BV in the spirit of (\ref{BonBV}).  

\vspace{0.1cm}

Let us mention that a direct proof of the conjecture for the off-shell action of the creation operator, as it was done for $N=1,2$ case, will need to fix the coefficients for the projection on the basis (\ref{basis}) given formally in remark \ref{proj}. Another possibility should be to give an indirect proof from the result of the SoV or from the recent development of the ODBA \cite{CYSW4}.  
Independently of the question of this proof, the MABA provide a constructive way to obtain the BV in a form that does not depend of the inhomogeneity parameters and solve the question of the homogeneous limit that is problematic in the SoV and not direct in the ODBA. 
Moreover the off-shell BV satisfy an off-shell equation for the action of the transfer matrix similar to the one for models with $U(1)$ symmetry. Thus, this off-shell equation appears to be a universal structure for quantum integrable models with our without $U(1)$ symmetry.
The off-shell criteria of the BV is of importance for other problems such as the construction of solutions of the quantum Knizhnik-Zamolodchikov \cite{RSV} or for the calculation of correlations functions in the ABA framework that can be reduced to the calculation of the scalar product between off-shell BV and an on-shell BV \cite{KMT,KKMNST}. In particular, the remarks \ref{SCBB} and \ref{SCBB2} give recursion relations for the scalar product in non-hermitian case.
This last question is actually considered in the case of the XXX chain on the segment \cite{BP}.  

\vspace{0.1cm}

Finally, let us remark that the proposed results for the Heisenberg XXZ spin-$\frac{1}{2}$  chain on the segment with triangular boundaries could allow one to study the thermodynamic limit $N\to \infty$ and must fit with the results obtained from the Onsager approach \cite{BB,BK2} that uses the vertex operator approach introduced by Jimbo {\it et al.} \cite{JKKKM}. 
In particular we must see if the result obtained by Cao {\it et al.} on the thermodynamic limit of the BE with an additional term \cite{termoXXZCao} can be applied to these specific cases and also if a determinant formula for the scalar product of the given BV can be obtained. For the latter, one has to construct the dual BV. For the former, we can remark that the two boundaries decouple up to order $N^{-2}$  \cite{termoXXZCao}  thus both cases should be equivalent if we don't care about finite size corrections.   
These results, following the spirit of \cite{KKMNST}, will allow one to obtain an alternative derivation of the integral representations of correlation functions and form factors given in \cite{BK2}. In addition, the ideas of the MABA and in particular the way of constructing the new operators could give some directions to extend the Onsager approach \cite{BB} to the case of generic boundary conditions.

\vspace{0.1cm}

{\bf Acknowledgement:} 
We thank N. Cramp\'e, R.A. Pimenta, R.I. Nepomechie for discussions and P. Baseilhac that suggest the adjective Modified in MABA to avoid confusion with usual and generalized ABA. We  thank J. Bajnok and the "MTA LendŸlet Holographic QFT Group, Wigner Research Centre for Physics" of Budapest for their invitation and hospitality during the workshop "Finite-size Technology in Low Dimensional Quantum Systems (VII)"  where a part of this work was done. We also thank the L2C of Montpellier where a part of this work was done. This work has been partially supported by Sao Paulo Research Foundation (FAPESP), grant 2014/09832-1.

\appendix 

\section{Functions and commutation relations\label{App:Func}}
We use the following functions 
\ben\label{fonctions1}
&&b(u)=\frac{u-u^{-1}}{q-q^{-1}}, \quad k^-(u) =\nu_-u+\nu_+u^{-1},\quad k^+(u) =\epsilon_+u+\epsilon_-u^{-1},\quad  c(u)=u^2-u^{-2}\\
&&\phi(u)= \frac{b(q^2u^2)}{b(qu^2)}, \quad m(u,v)=\frac{1}{b(u/v)b(quv)}, \quad F(u, v)=m(u,v)\frac{b(q^2 u^2)}{\phi(v)}, \\
&&f(u,v)= \frac{b(qv/u)b(uv)}{b(v/u)b(quv)}\ ,\quad g(u,v)= \frac{\phi(q^{-1}v^{-1})}{b(u/v)}, \quad w(u,v)= -\frac{1}{b(quv)},\\
 &&h(u,v)= \frac{b(q^2uv)b(qu/v)}{b(quv)b(u/v)},\quad k(u,v)= \frac{\phi(u)}{b(v/u)}, \quad
n(u,v)= \frac{\phi(u)\phi(q^{-1}v^{-1})}{b(quv)}\\
&&s(u,v)=\frac{\phi(q^{-1}u^{-1})}{b(v/u)b(q v^2)}, \quad x(u,v)=\frac{\phi(q^{-1}u^{-1})b(q u/v)}{b(u/v)b(q u v)}, \\
&& y(u,v)= -\frac{1}{b(q v^2)b(quv)}, \quad r(u,v) =\frac{\phi(q^{-1}u^{-1})}{b(v/u)}, \quad p(u,v)=\frac{b(u v)}{b(u/v)b(q u v)}. 
\een
Direct calculation gives the following relations 
\ben\label{idUWT1}
g(u,v)\phi(u)k^\pm(u)+n(u,v)k^\pm(q^{-1}u^{-1})&=&F(u,v)\phi(q^{-1} v^{-1})\phi(v)k^\pm(v),\\
\label{idUWT2}
k(u,v)k^\pm(q^{-1}u^{-1})+w(u,v)\phi(u)k^\pm(u)&=&-F(u,v)\phi(v)k^\pm(q^{-1}v^{-1}).
\een
From the RLL relation (\ref{RLL}), one can extract the commutations relations between the $l_{ij}$. Here, we will only need the following ones
\ben
&&l_{12}(u)l_{12}(v)=l_{12}(v)l_{12}(u),\label{coml12l12} \\
&&l_{11}(u)l_{12}(v)=\frac{b(qv/u)}{b(v/u)}l_{12}(v)l_{11}(u)+\frac{1}{b(u/v)}l_{12}(u)l_{11}(v),\label{coml11l12} \\
&&l_{22}(u)l_{12}(v)=\frac{b(qu/v)}{b(u/v)}l_{12}(v)l_{22}(u)+\frac{1}{b(v/u)}l_{12}(u)l_{22}(v).\label{coml22l12}\\
&&l_{21}(u)l_{12}(v)=l_{12}(v)l_{21}(u)+\frac{1}{b(u/v)}(l_{11}(u)l_{22}(v)-l_{11}(v)l_{22}(u)).\label{coml21l12}
\een
From the reflection algebra (\ref{RE}), one can extract  the commutations relations between the operators $\mathscr{A}$, $\mathscr{D}$, $\mathscr{C}$ and $\mathscr{B}$.
To order monomials of such operators in the basis span by operator valued series 
\ben
\mathscr{M}_{bdac}(\bar u,\bar v,\bar w,\bar x)=\mathscr{B}(\bar u)\mathscr{D}(\bar v)\mathscr{A}(\bar w)\mathscr{C}(\bar x)
\een
one needs the following commutation relations
\begin{eqnarray}
&&\mathscr{A}(u)\mathscr{B}(v)= f(u,v)\mathscr{B}(v)\mathscr{A}(u) + g(u,v)\mathscr{B}(u)\mathscr{A}(v) + w(u,v)\mathscr{B}(u)\mathscr{D}(v),\label{comAB}  \\
&&\mathscr{C}(v)\mathscr{A}(u)=f(u,v) \mathscr{A}(u)\mathscr{C}(v) +g(u,v)\mathscr{A}(v)\mathscr{C}(u) + w(u,v)\mathscr{D}(v)\mathscr{C}(u),\label{comCA}  \\
&&\mathscr{D}(u)\mathscr{B}(v)= h(u,v)\mathscr{B}(v) \mathscr{D}(u) +k(u,v)\mathscr{B}(u)\mathscr{D}(v)+ n(u,v)\mathscr{B}(u)\mathscr{A}(v),\label{comDB} \\
&&\mathscr{C}(u)\mathscr{D}(v)= h(u,v)\mathscr{D}(v) \mathscr{C}(u) +k(u,v)\mathscr{D}(u)\mathscr{C}(v)+n(u,v) \mathscr{A}(u)\mathscr{C}(v),\label{comCD}\\
&&\mathscr{C}(u)\mathscr{B}(v)=\mathscr{B}(v)\mathscr{C}(u)+s(u,v)\mathscr{A}(u)\mathscr{A}(v)+x(u,v)\mathscr{A}(v)\mathscr{A}(u)+y(u,v)\mathscr{D}(u)\mathscr{A}(v)\nonumber\\
&&\qquad\qquad\qquad\quad\quad+r(u,v)\mathscr{A}(u)\mathscr{D}(v)+p(u,v)\mathscr{A}(v)\mathscr{D}(u)+w(u,v)\mathscr{D}(u)\mathscr{D}(v),\label{comCB} \\
&&\mathscr{A}(u)\mathscr{D}(v)=\mathscr{D}(v)\mathscr{A}(u)+k(v,u)\big(\mathscr{B}(u)\mathscr{C}(v)-\mathscr{B}(v)\mathscr{C}(u)\big)\label{comAD}
\end{eqnarray}
and
\ben
&&\mathscr{A}(u)\mathscr{A}(v)=\mathscr{A}(v)\mathscr{A}(u)+w(u,v)\big(\mathscr{B}(u)\mathscr{C}(v)-\mathscr{B}(v)\mathscr{C}(u)\big)
,\label{comAA} \\
&&\mathscr{D}(u)\mathscr{D}(v)=\mathscr{D}(v)\mathscr{D}(u)-\phi(u)\phi(v)w(u,v)\big(\mathscr{B}(u)\mathscr{C}(v)-\mathscr{B}(v)\mathscr{C}(u)\big)
,\label{comDD} \\
&&\mathscr{B}(u)\mathscr{B}(v) = \mathscr{B}(v)\mathscr{B}(u),\label{comBB} \\
&&\mathscr{C}(u)\mathscr{C}(v) = \mathscr{C}(v)\mathscr{C}(u).\label{comCC} 
\end{eqnarray}
Let us remark that this set of relations is complete, {\it i.e.} they are isomorphic to the reflection equation.

\end{document}